\documentclass[a4paper,twocolumn,11pt,accepted=2023-04-10]{quantumarticle}
\pdfoutput=1
\usepackage[utf8]{inputenc}
\usepackage[english]{babel}
\usepackage[T1]{fontenc}
\usepackage{amsmath}
\usepackage{hyperref}

\usepackage{amsmath, empheq,braket,amsthm,amssymb,graphics,amsfonts,array,color,subfigure, mathrsfs}
\usepackage{latexsym}
\usepackage{tensor}
\usepackage{empheq}
\usepackage{bm}
\usepackage{xcolor}

\usepackage[numbers,sort&compress]{natbib}

\newtheorem{lemma}{Lemma}
\newtheorem{definition}{Definition}
\newtheorem{proposition}{Proposition}

\newcommand{\fr}{f_\mathrm{R}}
\newcommand{\gr}{g_\mathrm{R}}

\begin{document}

\title{Continuous majorization in quantum phase space}

\author{Zacharie Van Herstraeten}
\email{zvh@arizona.edu}
\affiliation{Centre for Quantum Information and Communication, \'Ecole polytechnique de Bruxelles, CP 165/59, Universit\'e libre de Bruxelles, 1050 Brussels, Belgium}
\affiliation{Wyant College of Optical Sciences, The University of Arizona, 1630 E. University Blvd., Tucson, AZ 85721, USA}
\author{Michael G. Jabbour}
\email{mgija@dtu.dk}
\affiliation{Centre for Quantum Information and Communication, \'Ecole polytechnique de Bruxelles, CP 165/59, Universit\'e libre de Bruxelles, 1050 Brussels, Belgium}
\affiliation{DAMTP, Centre for Mathematical Sciences, University of Cambridge, Cambridge CB3 0WA, United Kingdom}
\affiliation{Department of Physics, Technical University of Denmark, 2800 Kongens Lyngby, Denmark}
\author{Nicolas J. Cerf}
\email{ncerf@ulb.ac.be}
\affiliation{Centre for Quantum Information and Communication, \'Ecole polytechnique de Bruxelles, CP 165/59, Universit\'e libre de Bruxelles, 1050 Brussels, Belgium}

\begin{abstract}
\noindent We explore the role of majorization theory in quantum phase space. To this purpose, we restrict ourselves to quantum states with positive Wigner functions and show that the continuous version of majorization theory provides an elegant and very natural approach to exploring the information-theoretic properties of Wigner functions in phase space. After identifying all Gaussian pure states as equivalent in the precise sense of continuous majorization, which can be understood in light of Hudson's theorem, we conjecture a fundamental majorization relation: any positive Wigner function is majorized by the Wigner function of a Gaussian pure state (especially, the bosonic vacuum state or ground state of the harmonic oscillator). As a consequence, any Schur-concave function of the Wigner function is lower bounded by the value it takes for the vacuum state. This implies in turn that the Wigner entropy is lower bounded by its value for the vacuum state, while the converse is notably not true. Our main result is then to prove this fundamental majorization relation for a relevant subset of Wigner-positive quantum states which are mixtures of the three lowest eigenstates of the harmonic oscillator. Beyond that, the conjecture is also supported by numerical evidence. We conclude by discussing some implications of this conjecture in the context of entropic uncertainty relations in phase space.
\end{abstract}

\maketitle

\section{Introduction}
\label{sec:intro}

Majorization is an elegant and powerful algebraic theory that provides a means for comparing probability distributions in terms of disorder or randomness \cite{Hardy1934,Marshall2011}. It has been extensively employed during the last century in various fields such as mathematics \cite{Ando1989}, economics \cite{Mosler1994}, or information theory \cite{VanErven2010} where it can be used to derive inequalities for a variety of information-theoretic quantities such as entropies, see e.g. \cite{Alhejji2020,Jabbour2020} for recent works. Although its deep connections with unitary matrices had long been understood \cite{Horn}, it is only more recently that  majorization relations have been found to arise in quantum physics \cite{Nielsen1999}.
As such, it finds application in the study of entanglement transformations \cite{Nielsen1999,Nielsen2001}, in the discrimination of (distillable) entangled states \cite{Nielsen2001.2,Hiroshima2003}, in the derivation of quantum uncertainty relations \cite{Puchala2013,Rudnicki2014,Rudnicki2015}, or via so-called thermo-majorization in the context of quantum thermodynamics \cite{Brandao2015}, to cite a few. In the last years, majorization theory has also proven especially useful in the framework of bosonic quantum systems, which are our interest here, serving as an instrumental tool for the investigation of entropic inequalities that are paramount in the computation of the optimal communication rates of quantum communication systems \cite{Raul2012,Christos2013,DePalma2016,Jabbour2016,Jabbour2019}.

While majorization theory can be applied to both discrete and continuous probability distributions, the overwhelming majority of its applications in the existing literature concerns the former. In particular, while discrete majorization (the branch of the theory of majorization dealing with discrete probability distributions) has been the subject of numerous works in quantum information theory, notably including those dealing with bosonic quantum systems, continuous majorization (the branch concerned with continuous probability densities) has never been applied in this context.
Nevertheless, the phase-space formulation of quantum mechanics, which leads to a characterization of quantum states in terms of continuous distributions \cite{Leonhardt2010}, hints at the great potential of continuous majorization in such a framework. This is especially highlighted by the strong connection between majorization and entropies (both classical and quantum) coupled with the great interest directed towards continuous entropic uncertainty relations in recent years \cite{Hertz2017,Hertz2019}.

In the present paper, we argue and demonstrate that the theory of continuous majorization is highly relevant in the context of quantum physics. It provides a perfect solution in order to compare quantum phase-space distributions in terms of intrinsic disorder. It then also provides a natural means to address entropic properties in phase space, hence suggesting a fresh new perspective to entropic uncertainty relations.
Consider a quantum system characterized by the usual pair of canonically-conjugate continuous variables $x$ and $p$.
A pure state of the system is described by the complex wave functions $\psi(x)$ or $\varphi(p)$, which are related by a Fourier transform. The operators $\hat{x}$ and $\hat{p}$ associated with the observables $x$ and $p$ obey the canonical commutation relation $\left[\hat{x},\hat{p}\right]=i \hbar$. The observables $x$ and $p$ may designate the position and momentum variables, or also for instance two canonically-conjugate quadratures of the electromagnetic field (we will adopt this quantum optics language in the rest of this paper although our results and conclusions hold true in general for any canonical pair). The phase-space representation of a pure state is embodied by its Wigner function, which is a two-dimensional continuous function defined as \cite{Leonhardt2010}
\begin{equation}
	W(x,p) = 
	\dfrac{1}{\pi \hbar}
	\int
	\mathrm{e}^{2 i p y / \hbar}
	\psi^{\ast}\left(x+y\right)
	\psi\left(x-y\right)
	\mathrm{d}y .
\label{eq:def_Wigner_func_pure}	
\end{equation}
The Wigner function can in many ways be thought of as a joint probability distribution of the variables $x$ and $p$. For instance, the probability densities of $x$ and $p$, respectively $\rho_x(x)=\vert\psi(x)\vert^2$ and $\rho_p(p)=\vert\varphi(p)\vert^2$, can be retrieved from the marginal distributions of the Wigner function as $\rho_x(x) = \int W(x,p)\, \mathrm{d}p$ and $\rho_p(p) = \int W(x,p)\, \mathrm{d}x$. However, the non-commutativity of operators $\hat{x}$ and $\hat{p}$ has deep implications regarding the existence of such a joint distribution as quantum mechanics forbids simultaneously fixing the values of two non-commuting variables. In the phase-space description of quantum states, this translates into the fact that Wigner functions may take negative values in general, making them so-called \textit{quasi-probability} distributions.

The necessary existence of negative Wigner functions in quantum mechanics can also be justified with a simple argument relying on the overlap formula, which reads \cite{Leonhardt2010}
\begin{equation}
	\vert
	\langle
	\psi_1
	\vert
	\psi_2
	\rangle
	\vert^2
	=
	2\pi\iint
	W_{1}(x,p) \, W_{2}(x,p) \,
	\mathrm{d}x \, \mathrm{d}p  ,
	\label{eq:overlap}
\end{equation}
where $W_1$ and $W_2$ are the Wigner functions associated respectively with $\psi_1$ and $\psi_2$.
Having in mind that there exist pure states with Wigner functions positive everywhere, namely Gaussian states (\textit{i.e.}, states whose Wigner function is a Gaussian distribution \cite{Weedbrook2012}), it follows that any pure state that is orthogonal to a Gaussian pure state must have a Wigner function that takes negative values. In what follows, we will qualify a state with a positive Wigner function as \textit{Wigner-positive} and refer to the corresponding property as \textit{Wigner-positivity} \cite{VanHerstraeten2020}. Note that Wigner-positivity is a particular case of $\eta$-positivity for $\eta=0$ \cite{Narcowich1989, Brocker1995}. We will restrict our analysis in this paper to Wigner-positive states and explore whether majorization theory can be applied to their Wigner functions. In fact, the set of \textit{pure} Wigner-positive states is well understood: Hudson's theorem \cite{Hudson1974, soto1983wigner} establishes that a pure state is Wigner-positive if and only if it is a Gaussian pure state. Furthermore, all Gaussian pure states happen to be related to each other through Gaussian unitaries in state space, which are associated with symplectic transformations in phase space \cite{Weedbrook2012}.  For a single mode, these affine transformations are simply combinations of displacements, rotations and squeezings of the Wigner function. Crucially for what follows, these symplectic transformations have the property that they \textit{preserve areas} in phase space.

The concept of area in phase space can be related to the notion of \textit{level-function} \cite{Marshall2011}, which happens to be at the core of the theory of continuous majorization. For a given (positive) Wigner function $W(x,p)$, the level-function associates to a value $t$ the area of the region in phase space that has $W(x,p)$ greater than $t$.
In what follows, any two (positive) Wigner functions that have the same level-functions will be called \textit{level-equivalent}.
Area-preserving transformations are transformations that keep the level-function unchanged, hence the Wigner functions of all Gaussian pure states are level-equivalent. This leads to the following property, which can be viewed as a corollary of Hudson's theorem: 
\begin{quote}
	Any \textit{pure} Wigner-positive state has a Wigner function that is \textit{level-equivalent} to that of a Gaussian pure state.
\end{quote}

\noindent Note that all level-equivalent functions to the Wigner function of a Gaussian pure state do not necessarily describe a \textit{physical} Wigner function (that is, corresponding to a positive semidefinite density operator); but if it is physical, then we know that it corresponds to a Wigner-positive pure state according to Hudson's theorem.

This begs the question whether the above approach building on level-functions can be generalized to mixed quantum states. 
A mixed state $\hat{\rho} = \sum p_i\ket{\psi_i}\bra{\psi_i}$ is a statistical mixture of pure states and its Wigner function $W=\sum p_i W_i$ is the corresponding mixture of their Wigner functions. It is straightforward to see that there exist non-Gaussian mixed states characterized by positive Wigner functions, a simple way to construct such a state being to form a non-Gaussian mixture of Gaussian pure states. A natural question is then whether one can characterize the full set of Wigner-positive mixed states. Such a problem happens to be difficult and remains only very partially solved today \cite{narcowich1986necessary,Brocker1995,Mandilara2009,mandilara2010gaussianity}, in the sense that no satisfying extension of Hudson's theorem to mixed states has been stated in the literature.

In the present work, we apply the theory of majorization to positive Wigner functions in order to tackle the aforementioned problem. While we do not provide a complete description of the full set of Wigner-positive states, we offer a way to compare such states in terms of continuous majorization. Continuous majorization generalizes the concept of level-equivalent functions by allowing one to compare functions that are not level-equivalent.
Intuitively, for two functions $f$ and $g$, the relation $f$ majorizes $g$ (in the continuous sense) means that $f$ is more narrow (i.e., more ordered) than $g$, while two functions that are level-equivalent majorize each others. This leads us to conjecture the following generalized statement:
\begin{quote}
	Any \textit{mixed} Wigner-positive state has a Wigner function \textit{majorized} by that of a Gaussian pure state.
\end{quote}
\noindent While the above statement may seem natural as one would expect a mixed state to be more disordered than a pure state, it is important to stress that the notion of disorder in the statement refers to distributions in phase space. It cannot be related in any simple way to disorder in state space, which concerns density operators. As a consequence, proving this conjecture is far from straightforward.

The objective of this paper is to develop the theory of continuous majorization in the framework of quantum physics and then prove the above statement for some carefully chosen positive Wigner functions. In Section \ref{sec:cont_major}, we begin by introducing the theory of continuous majorization, as this is not a standard topic for physicists. This then provides us with the proper mathematical tools to apply continuous majorization to quantum phase space in Section \ref{sec:results}, before giving a proof of our conjecture restricted to a subset of Wigner-positive states. Finally, we discuss our results in connection with the notion of Wigner entropy \cite{VanHerstraeten2020} and apply them in the context of entropic uncertainty relations in Section \ref{sec:conclusion}.

\section{Continuous majorization}
\label{sec:cont_major}

We are now going to lay out the basics of majorization theory in the context of continuous probability distributions, giving rigorous definitions for the concepts mentioned in the introduction. We consider $n$-dimensional continuous non-negative and integrable distributions on $\mathbb{R}^n$ or $\mathbb{R}_+$, depending on the situation. The distribution $f : \mathcal{A} \rightarrow \mathbb{R}_+$, where $\mathcal{A}$ can be  $\mathbb{R}^n$ (for any positive integer $n$) or $\mathbb{R}_+$, is a genuine probability distribution if
\begin{equation} \label{eq:probDist}
	f\left(\mathbf{r}\right)\geq 0\quad\forall \mathbf{r}\in\mathcal{A}
	,
	\qquad
	\int_{\mathcal{A}}
	f(\mathbf{r}) \, 
	\mathrm{d}\mathbf{r}
	=
	1.
\end{equation}
Hereafter, we omit the bounds in integrals as long as the integration is performed on the whole domain $\mathcal{A}$, as in the normalization identity in Eq.~\eqref{eq:probDist}. Note that the distributions are normalized to $1$ but the definitions and properties we show in what follows still hold if the normalization constant is different (provided it is the same for all functions we consider). Furthermore, while we consider functions $f$ defined on an infinite domain $\mathcal{A}$ for our purposes here, in which case they have to be non-negative everywhere, note that majorization can also be applied to partially negative functions if they are defined on a finite domain \cite{Marshall2011}. 

As we mentioned in the introduction, a core element of the theory of majorization is the level-function, which we now rigorously define.
\begin{definition}
The level-function $m_f : \mathbb{R}_+ \rightarrow \mathbb{R}_+$ of a distribution $f : \mathcal{A} \rightarrow \mathbb{R}_+$ is the function
\begin{equation}
m_f(t) = \nu\left(\lbrace \mathbf{r} \in \mathcal{A} :f(\mathbf{r})\geq t\rbrace\right),
\end{equation}
where $\nu$ stands for the Lebesgue measure.
\end{definition}

\begin{figure}	\includegraphics[width=1\linewidth]{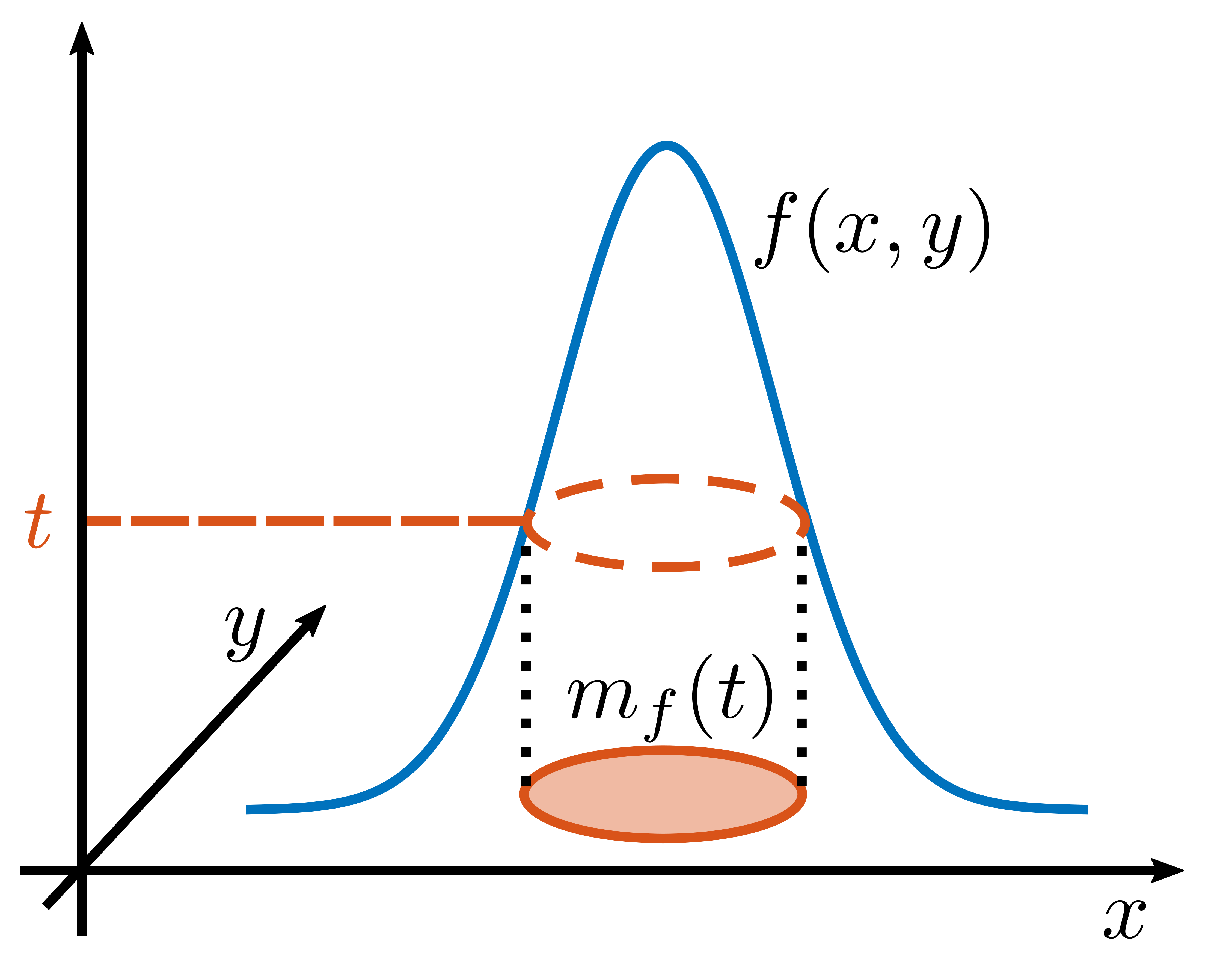}
	\caption{Illustration of the level-function of a two-dimensional distribution. The distribution $f(x,y)$ is defined over the plane $(x,y)$. The value of the level-function $m_f(t)$ is the area of the domain (in red) such that $f(x,y)\geq t$.}
	\label{fig:level_function_2d}
\end{figure}

\noindent Basically, the level-function $m_f$ is the size of the subdomain of $f$ that contains elements whose corresponding image under $f$ is higher than $t$. Figure~\ref{fig:level_function_2d} provides an example of the level-function of a two-dimensional distribution. As we are going to show, all  information needed to compare two distributions using a majorization relation (from here onwards, we write majorization for continuous majorization) is enclosed in their level-functions.
When $f$ and $g$ are level-equivalent, meaning they have the same level-functions $m_f(t) = m_g(t)$ for all $t$, we use the notation $f \equiv g$. While level-equivalent distributions can have very different shapes, they nonetheless are comparable in various ways. In order to see this, consider any function $\varphi$ defined on $\mathbb{R}_+$ and obeying some general conditions  (see \cite{Wang2014}), which can be turned into a functional $\phi$ acting on the probability distribution $f$ as  
\begin{equation}
	\phi(f) = \int \varphi\left(f(\mathbf{r})\right) \,  \mathrm{d}\mathbf{r}.
	\label{eq:phi_int_of_phi}
\end{equation}
Since the integral is carried out over the whole domain, $\phi$ is invariant when parts of the domain are ``swapped". It then only depends on the ``size'' of the domain associated to each value taken by the distribution $f$. This is precisely the information carried by the level-function, which simply corresponds for each value of $t$ to the choice $\varphi(x)=\Theta (x-t)$ with $\Theta$ being the Heaviside step function, so that $\phi$ is the same for two level-equivalent distributions, that is, $\phi(f) = \phi(g)$ if $f \equiv g$. Note that for the choice of the function $\varphi(x)=-x \ln x$, we get the functional
\begin{equation}
	\phi(f) = - \int f(\mathbf{r}) \ln f(\mathbf{r}) \,  \mathrm{d}\mathbf{r}  ,
	\label{eq:Shannon_entropy}
\end{equation}
which is nothing else but Shannon's differential entropy of the probability distribution $f$. Therefore, two level-equivalent distributions have the same Shannon entropy; they are comparable in their randomness content.

\begin{figure}[t]
\includegraphics[width=1\linewidth]{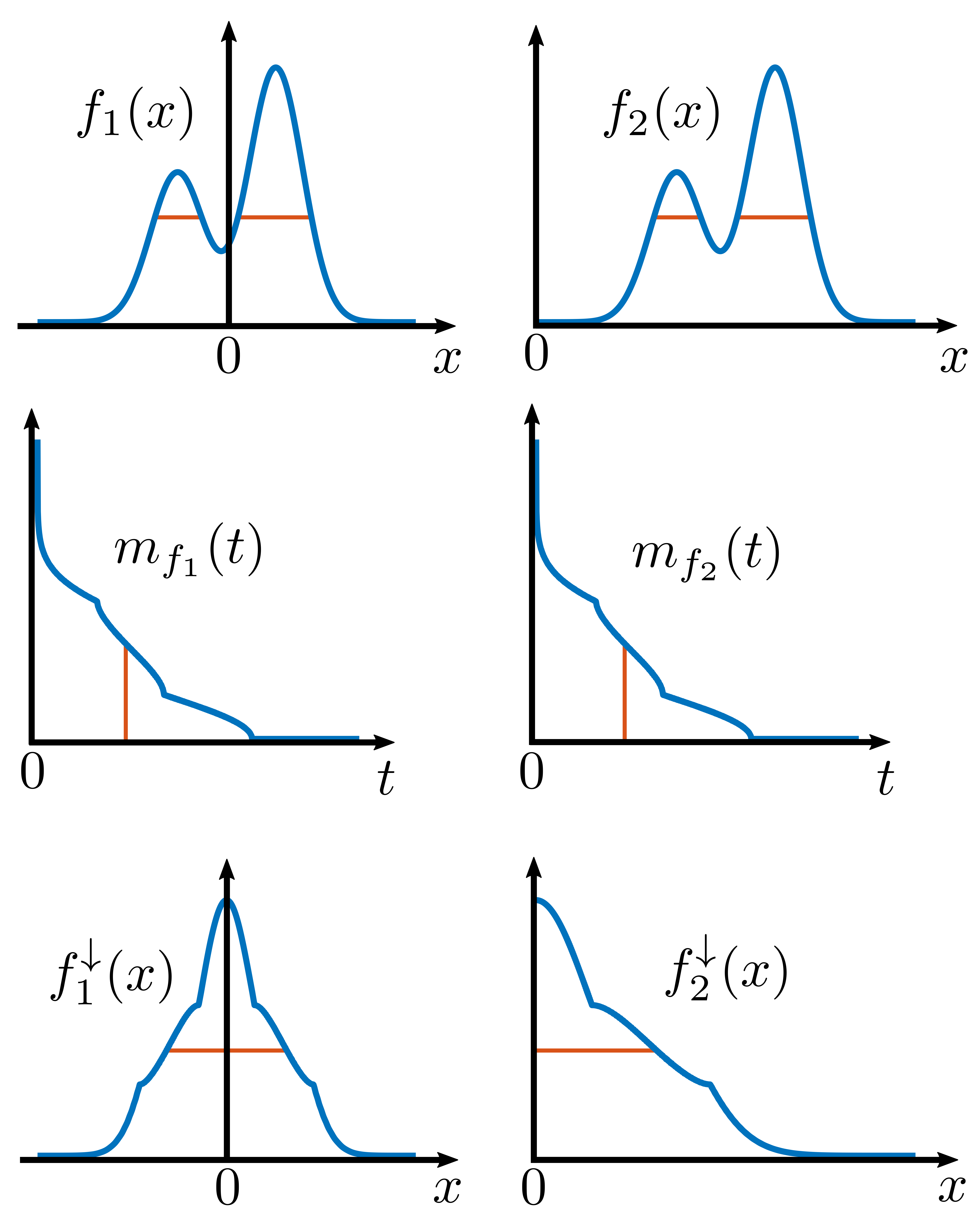}
	\caption{
		Schematic of the decreasing rearrangement. In the first row, $f_1$ and $f_2$ are two examples of one-dimensional probability distributions respectively defined on $\mathbb{R}$ and $\mathbb{R}_+$. They both have the same level-functions $m_{f_1}$ and $m_{f_2}$, which are pictured in the second row. 
		They, however, have different decreasing rearrangements $f_1^{\downarrow}$ and $f_2^{\downarrow}$, as shown on the last row. The red segments represent the domain that corresponds to a given value of $t$ and therefore have the same length in the different plots. Note that for a distribution defined on $\mathbb{R}_+$ such as $f_2$, the decreasing rearrangement $f_2^{\downarrow}$ is simply the inverse function of the level-function $m_{f_2}$.}
	\label{fig:dec_rear}
\end{figure}

For a given level-function $m_f$, one can build an infinite number of level-equivalent distributions whose level-functions are given by $m_f$. Among the set of distributions sharing the same level-function, the so-called \textit{decreasing rearrangement} plays a prominent role in majorization theory. It is defined as follows \cite{Marshall2011}.
\begin{definition} \label{def:deacReag1}
	The decreasing rearrangement $f^\downarrow$ of a function $f$ defined on a domain $\mathcal{A}$ is the unique function defined on the same domain $\mathcal{A}$ that is radial-decreasing and level-equivalent to $f$.
\end{definition}
\noindent An $n$-dimensional distribution $f$ is radial if it can be written as a function that only depends on the norm of its argument, \textit{i.e.}, $f(\mathbf{r})=\fr(\Vert\mathbf{r}\Vert)$, with $\fr:\mathbb{R}_+ \rightarrow \mathbb{R}_+$. Furthermore, the function $f$ is said to be radial decreasing if $\fr(r_1) \geq \fr(r_2)$ when $r_1 < r_2$.
Note here that the term decreasing is used instead of non-increasing when referring to such rearrangements in the literature.
The decreasing rearrangement $f^\downarrow$ of a distribution $f$ takes its maximum value at the origin and decreases monotonically as it gets farther from it. It is easy to understand that it only depends on the level-function $m_f$ of $f$. Moreover, $f^\downarrow$ is the same for any distribution that is level-equivalent to $f$, as it is unique. Examples of decreasing rearrangements are pictured in the lower part of Fig.~\ref{fig:dec_rear}.

We are now in position to define the majorization relation between two distributions.
\begin{definition}[Continuous majorization \cite{Hardy1929,Joe1987}]
Let $f$ and $g$ be two probability distributions defined on the same domain $\mathcal{A}$.
The distribution $f$ majorizes the distribution $g$, written $f\succ g$, if and only if
\begin{equation}
\int\limits_{\Vert \mathbf{r}\Vert\leq s} f^{\downarrow}(\mathbf{r})  \,   \mathrm{d}\mathbf{r}
\quad
\geq
\quad
\int\limits_{\Vert \mathbf{r}\Vert\leq s} g^{\downarrow}(\mathbf{r})  \,   \mathrm{d}\mathbf{r}
\qquad
\forall \, s \geq 0, 
\label{eq:def_major}
\end{equation}
with equality when $s$ tends to infinity. 
\label{def:def_major}
\end{definition}
\noindent The equality when $s$ tends to infinity imposes that $f$ and $g$ be normalized to the same value. When the inequalities in Eq. \eqref{eq:def_major} are reversed, we say that $f$ is majorized by $g$, written $f\prec g$. If both $f \succ g$ and $f \prec g$ hold, then $f$ and $g$ are level-equivalent and the inequalities in Eq. \eqref{eq:def_major} become equalities. If neither $f\succ g$ nor $f\prec g$ holds, we say that $f$ and $g$ are incomparable.

It is useful in practice to define the objects appearing on both sides of the inequalities in Eq.~\eqref{eq:def_major}, mainly for ease of notations. The \textit{cumulative integral} of a distribution $f$ is the function $S_f : \mathbb{R}_+ \rightarrow \mathbb{R}_+$ defined as
\begin{equation}
S_f(s) 
= 
\int\limits_{\Vert \mathbf{r}\Vert\leq s} f^{\downarrow}(\mathbf{r})  \,   \mathrm{d}\mathbf{r}.
\label{eq:cumulative_integral}
\end{equation}
It can be understood as the highest value that can be obtained by integrating any level-equivalent function to $f$ over a ball of radius $s$. Equivalently, it is the highest value of the integral of $f$ over any part of the domain (possibly made of non-contiguous regions) that has a volume equal to that of a ball of radius $s$. The majorization relation $f \succ g$ then holds if and only if $S_f(s) \geq S_g(s)$ for all $s \geq 0$, with equality when $s$ tends to infinity.

Since Definition \ref{def:def_major} refers to the decreasing rearrangements $f^{\downarrow}$ and $g^{\downarrow}$, which are defined based on the level-functions $m_f$ and $m_g$, it is clear that the majorization relation $f \succ g$ solely depends on the level-functions of $f$ and $g$. Unlike in the discrete case, the decreasing rearrangement of a function is often hard to compute, which makes Definition \ref{def:def_major} difficult to use. There are, however, equivalent statements that are less cumbersome. We introduce two such statements hereafter as Propositions \ref{prop:major_equiv_plus} and \ref{prop:convex_equiv}. We point the interested reader to References \cite{Marshall2011,Wang2014} for proofs and details of these propositions.
\begin{proposition}
	Let $f$ and $g$ be two probability distributions defined on the same domain $\mathcal{A}$.
	We have that $f \succ g$ if and only if
	\begin{equation}
	\int 
	\big[
	f(\mathbf{r})-t
	\big]_+
	\mathrm{d}\mathbf{r}
	\geq
	\int 
	\big[
	g(\mathbf{r})-t
	\big]_+
	\mathrm{d}\mathbf{r}
	\qquad\forall \, t \geq 0,
	\label{eq:major_equiv_plus}
	\end{equation}
	with equality when $t=0$.
	\label{prop:major_equiv_plus}
\end{proposition}
\noindent The notation $\left[\hspace{1pt}\cdot\hspace{1pt}\right]_+$ is such that $\left[z\right]_+=z$ if $z\geq 0$ and $\left[z\right]_+=0$ otherwise.
Note that the function $x \mapsto \left[\hspace{1pt}x - t \hspace{1pt}\right]_+$ acting on $f(\mathbf{r})$ and $g(\mathbf{r})$ in Eq. \eqref{eq:major_equiv_plus} can be viewed as a special case of the function $\varphi(x)$ used in Eq.~\eqref{eq:phi_int_of_phi}. Hence, Proposition \ref{prop:major_equiv_plus} is again solely characterized by the level-functions of $f$ and $g$.


Proposition \ref{prop:major_equiv_plus} is useful to prove a property that we will later need, namely that if $f\succ g_1$ and $f\succ g_2$, then $f$ majorizes any convex combination of $g_1$ and $g_2$, \textit{i.e.},
\begin{equation}
	\begin{cases}
		f\succ g_1
		\\
		f\succ g_2
	\end{cases}
	\Rightarrow
	\quad
	f\succ \lambda g_1 + (1-\lambda)g_2  ,
\label{eq:majorization_convex_mixture}	
\end{equation}
with $0\leq\lambda\leq 1$. Using the fact that the function $\gamma_t(z) = \left[z-t\right]_+$ is convex in $z$ for all $t\geq 0$ and using Jensen's inequality, we have
$\gamma_t\Big(\lambda \, g_1(\mathbf{r})+(1-\lambda)\, g_2(\mathbf{r})\Big) \le \lambda \,\gamma_t(g_1(\mathbf{r})) +(1-\lambda)\,\gamma_t(g_2(\mathbf{r}))$. Equation \eqref{eq:majorization_convex_mixture}	then follows from integrating this inequality over $\bf r$ and using Proposition \ref{prop:major_equiv_plus}.

\begin{proposition}
	Consider two probability distributions $f$ and $g$ defined on the same domain $\mathcal{A}$. We have that $f \succ g$ if and only if
	\begin{equation} \label{eq:prop:convex_equiv}
	\int \varphi\left(f(\mathbf{r})\right)\mathrm{d}\mathbf{r}
	\geq
	\int \varphi\left(g(\mathbf{r})\right)\mathrm{d}\mathbf{r}
	\end{equation}
	holds for all continuous convex functions $\varphi :\mathbb{R}_+\rightarrow\mathbb{R}$ with $\varphi(0)=0$, for which  the integrals exist on both sides (see \cite{Marshall2011}, p. 607).
	\label{prop:convex_equiv}
\end{proposition}
\noindent Proposition \ref{prop:convex_equiv} is particularly useful as Eq. \eqref{eq:prop:convex_equiv} implies inequalities on quantities that are functionals of $f$ and $g$ written in the form of  Eq. \eqref{eq:phi_int_of_phi}. Thus, for any such functional $\phi$, the majorization relation $f \succ g$ implies that $\phi(f)\ge \phi(g)$. 
One such functional is (up to a sign) the Shannon's differential entropy, Eq. \eqref{eq:Shannon_entropy}, which is traditionally denoted $h(f)$. As a consequence, if $f \succ g$, then $h(f)\le h(g)$, which is consistent with the idea that $f$ is more ordered than $g$, so it has a lower entropy.

As a matter of fact, the inequalities \eqref{eq:prop:convex_equiv} hold for an even more general set of functionals beyond the special form \eqref{eq:phi_int_of_phi}, namely the so-called Schur-convex functions (we call them functions here, instead of functionals). This is actually the definition of the set of Schur-convex functions \cite{Schur1923}, which have been defined for discrete probability distributions through a discrete majorization relation but can be equivalently defined in the continuous case as follows \cite{Marshall2011}:
\begin{definition}[Schur-convex functions]
	A real-valued function $\phi$ defined on the set of probability distributions on $\mathcal{A}$ is called Schur-convex if, for any pair of probability distributions $f$ and $g$ defined on $\mathcal{A}$, we have
	\begin{equation}
		f\succ g \quad \Rightarrow \quad \phi(f)\geq\phi(g).
	\label{eq:Shur_concave}		
	\end{equation}
	The function $\phi$ is said to be Schur-concave if the opposite inequality holds.
\label{def:Shur_concave}	
\end{definition}
\noindent It is easy to see that Schur-convex functions take the same value for level-equivalent distributions. Furthermore, any real-valued function $\phi$ defined on the set of probability distributions on $\mathcal{A}$ that is convex and takes the same values for level-equivalent distributions in $\mathcal{A}$ is necessarily Schur-convex \cite{Roberts1973}. For instance, any functional $\phi$ of the form of \eqref{eq:phi_int_of_phi} is Schur-convex, provided that $\varphi$ is convex. The opposite, however, is not necessarily true. This is illustrated, for instance, with the R\'enyi entropy \cite{Renyi1961},
\begin{equation}
	h_{\alpha}\left(f\right)
	=
	\dfrac{1}{1-\alpha}
	\ln\left(
	\int f\left(\mathbf{r}\right)^\alpha\mathrm{d}\mathbf{r}
	\right),
\end{equation}
where the parameter $\alpha$ can be chosen such that $0<\alpha<1$ or $\alpha>1$. Indeed, R\'enyi entropies can be proven to be concave only for $\alpha \in [0,1)$, while 
they are always Schur-concave \cite{He2003}. In the limit where $\alpha \rightarrow 1$, the R\'enyi entropy reduces to Shannon's differential entropy, that is, $h_{\alpha\to 1}\left(f\right)=h(f)$. 
Regardless of the value of  $\alpha$, the R\'enyi entropies can thus always be associated to some measure of  disorder, in the sense that  if $f \succ g$, then $h_{\alpha}(f)\le h_{\alpha}(g)$, as a consequence of the Schur-concavity of $h_{\alpha}$.
More generally, all Schur-concave functions $\phi$  can be understood in view of Definition \ref{def:Shur_concave} as generalized measures of disorder that are consistent with majorization theory: if $f$ is more ordered than $g$ as expressed by $f \succ g$, then it has a lower measure of disorder $\phi(f)\le \phi(g)$.

To be complete, we mention another characterization of continuous majorization that is based on semidoubly stochastic operators, in analogy with the characterization of discrete majorization between infinite probability vectors in terms of semidoubly stochastic matrices 
\begin{proposition}
	Let $f$ and $g$ be two probability distributions defined over the same domain $\mathcal{A}$.
	Let $f$ and $g$ be related via
	\begin{equation}
		g(\mathbf{r})
		=
		\int B(\mathbf{r},\mathbf{s})\, f(\mathbf{s})\, \mathrm{d}\mathbf{s},
  \label{eq-condition-proposition3}
	\end{equation}
	where $B:\mathcal{A} \times \mathcal{A} \mapsto\mathbb{R}$ is the kernel of some semidoubly stochastic operator, \textit{i.e.}, $B(\mathbf{r},\mathbf{s})\geq 0$, $ \forall \mathbf{r},\mathbf{s}$, $\int B(\mathbf{r},\mathbf{s})\, \mathrm{d}\mathbf{r}=1$, $\forall\mathbf{s}$, and $\int B(\mathbf{r},\mathbf{s})\, \mathrm{d}\mathbf{s}\leq 1$, $\forall\mathbf{r}$. Then $f\succ g$.
\label{prop:semi-doubly-stochastic}
\end{proposition}
In contrast with Propositions \ref{prop:major_equiv_plus} and \ref{prop:convex_equiv}, Proposition \ref{prop:semi-doubly-stochastic} only provides a sufficient condition for majorization. 
 Note that if the probability distributions were defined over a finite-size domain $\mathcal{A}$, condition \eqref{eq-condition-proposition3} has then been proven to be necessary as well \cite{Ryff1965-mk}.
However, to our knowledge, there is no proof of the equivalence between the existence of a semidoubly stochastic operator and a majorization relation for distributions defined over an infinite-size continuous domain (although it is plausible that it holds when the domain is $\mathbb{R}^n$). This is a current topic
of research in mathematics, see Refs. \cite{bahrami2021semi, manjegani2023majorization}.



In order to prove our main result in Sec. \ref{sec:results}, we will actually use another sufficient condition for majorization (see Lemma \ref{lem:mixture_of_distributions}), which gives a very clear interpretation of the meaning of  $f \succ g$ .

\section{Phase-space majorization}
\label{sec:results}

We are now in position to apply majorization theory in quantum phase space and formulate our main conjecture in proper mathematical terms. We consider a single-mode bosonic system modelled by a quantum harmonic oscillator, e.g., a mode of the electromagnetic field. For the sake of simplicity, we take the convention $\hbar = 1$, so the observables $\hat{x}$ and $\hat{p}$ obey the commutation relation $\left[\hat{x},\hat{p}\right]=i$. By defining the annihilation operator $\hat{a} = (\hat{x}+i\hat{p})/\sqrt{2}$, the creation operator $\hat{a}^\dagger = (\hat{x}-i\hat{p})/\sqrt{2}$, and the number operator $\hat{n} = \hat{a}^\dagger\hat{a}$, the Hamiltonian of the system takes the simple form $\hat{H}=\hat{n}+1/2$. 
Its eigenstates are the Fock states, denoted as $\ket{n}$ for $n = 0, 1, \cdots$, and associated with the wave functions \cite{Leonhardt2010}
\begin{align}
	\psi_n(x) &= \pi^{-\frac{1}{4}}2^{-\frac{n}{2}}\left(n!\right)^{-\frac{1}{2}}H_n(x)\exp\left(-\frac{x^2}{2}\right),
	\label{eq:wave_function_fock}
\end{align}
where $H_n$ are Hermite polynomials. Any state $\rho$ of the harmonic oscillator can be associated with its Wigner function \cite{Leonhardt2010}
\begin{equation}
	W(x,p) = 
	\dfrac{1}{\pi}
	\int
	\mathrm{e}^{2 i p y } \langle x-y | \rho | x+y \rangle
	\mathrm{d}y ,
	\label{eq:def_Wigner_func_mixed} 
\end{equation}
which reduces to Eq. \eqref{eq:def_Wigner_func_pure} if $\rho$ denotes a pure state. In particular, the Fock states $\ket{n}$ are associated with the Wigner functions \cite{Leonhardt2010}
\begin{align}
	W_n(x,p) &= 
	\frac{(-1)^n}{\pi} L_n\left(2x^2+2p^2\right)\exp\left(-x^2-p^2\right),
	\label{eq:wigner_function_fock}
\end{align}
where $L_n$ are Laguerre polynomials. Note that $W_n(x,p)$ exhibit a rotational symmetry, making Fock states \textit{phase-invariant}. In fact, any phase-invariant state can be expressed as a mixture of Fock states. As we shall see, the Fock state associated with $n=0$ (i.e., the vacuum state) plays a key role with regard to majorization. It admits the Wigner function 
\begin{equation} \label{eq:W0}
	W_0(x,p) = \dfrac{1}{\pi}\exp\left(-x^2-p^2\right),
\end{equation}
which we write in short as $W_0(r)=\exp\left(-r^2\right)/\pi$, using the non-negative parameter $r$ such that $r^2=x^2+p^2$ [note the slight abuse of notation as $W_0$ is used both as a function of $r$ and $(x,p)$ hereafter]. According to Hudson's theorem, it is the only pure state admitting a positive Wigner function (up to symplectic transformations).

Frow now on, let us restrict to Wigner-positive states and denote the set of Wigner functions that are positive everywhere as $\mathcal{W}_+$. Clearly, all Wigner functions in $\mathcal{W}_+$ are genuine probability distributions, so that a question that arises naturally is whether the majorization relation introduced in Section \ref{sec:cont_major} has a meaning when applied to (positive) Wigner functions. A first clue that this may be the case follows from the Wigner entropy \cite{VanHerstraeten2020} of a quantum Wigner-positive state, which is the Shannon differential entropy of the Wigner function associated with the state, denoted as $h(W)$. In Ref. \cite{VanHerstraeten2020}, it is conjectured (and proved in some special cases) that 
\begin{equation}
h(W)\ge h(W_0)  \qquad
	\forall \, W \in \mathcal{W}_+  . 
\label{eq:conjecture_Wigner_entropy}	
\end{equation}
Since $h(W)$ can be understood as a special case of a functional $\phi$ of the Wigner function $W$, as in Eqs. \eqref{eq:phi_int_of_phi}
and \eqref{eq:Shannon_entropy}, it is striking to conjecture that Eq. \eqref{eq:conjecture_Wigner_entropy} is a consequence of a fundamental majorization relation, 
following the lines of Eq. \eqref{eq:Shur_concave}. This is what we do now.

\subsection*{Majorization conjecture}

Let us denote as $\mathcal{W}_+^\mathrm{pure}$ the set of pure Wigner-positive states, which is a subset of $\mathcal{W}_+$. As a consequence of Hudson's theorem,  $\mathcal{W}_+^\mathrm{pure}$ only contains Gaussian pure states and actually contains all of them.  Remarkably, the Wigner functions of all Gaussian pure states are level-equivalent since they are all related by symplectic transformations in phase space. Indeed, a symplectic transformation is an affine linear map on the vector $\mathbf{r}=(x,p)^T$, namely $\mathbf{r} \mapsto  \mathbf{r'} \equiv S \, \mathbf{r} + \mathbf{d}$ where $S$ is a symplectic matrix and $\mathbf{d}$ is a displacement vector, see \cite{Weedbrook2012} for mode details. By denoting the Wigner function before and after the symplectic transformations as $W(\mathbf{r})$ and $W'(\mathbf{r'})$, respectively, we have $W'(\mathbf{r'})= W(\mathbf{r}) / |\det S|$.  By doing the change of variables, the  level-functions corresponding to $W$ and $W'$ can then be shown to coincide as
\begin{eqnarray}
m_{W'}(t) &=& \int  \Theta \Big(  W'(\mathbf{r'}) - t  \Big)  \, \mathrm{d}\mathbf{r'}   \nonumber \\
&=&  \int  \Theta \Big(  W(\mathbf{r}) / |\det S| - t  \Big) \, |\det S| \, \mathrm{d}\mathbf{r}   \nonumber  \\
&=&  \int  \Theta \Big(  W(\mathbf{r}) - t  \Big)  \, \mathrm{d}\mathbf{r}    \nonumber  \\
&=& m_{W}(t)
\end{eqnarray}
where we have used the fact that $\det S=1$ for a symplectic transformation.

Hence, all Gaussian pure states have a Wigner function that is level-equivalent to $W_0$, making them all equivalent to $W_0$ from the point of view of majorization, namely
\begin{equation}
	W_0 \equiv W,
	\qquad
	\forall \, W \in \mathcal{W}_+^\mathrm{pure}.
\end{equation}
With this in mind, our main conjecture can be restated as follows:
\begin{equation} \label{eq:conjW0majW+}
	W_0 \succ W,
	\qquad
	\forall \, W \in \mathcal{W}_+.
\end{equation}
This expresses that, in the sense of majorization theory,  the most fundamental (Wigner-positive) state is the vacuum state, {\it  i.e.}, the ground state of the Hamiltonian of the harmonic oscillator.
Note that conjecture \eqref{eq:conjW0majW+} goes beyond the scope of quantum optical states and applies to the phase space associated with any canonical pair $(x,p)$. Furthermore, it is unrelated to the Hamiltonian of the system : the (positive) Wigner function of any state of the system must always be majorized by function \eqref{eq:W0}. As discussed in Sec. \ref{sec:conclusion}, function \eqref{eq:W0} can therefore be associated with the lowest-uncertainty state, even if it entails the lowest energy for the harmonic oscillator only.

\subsection*{Restricted proof }

\begin{figure}
    \includegraphics[width=\linewidth]{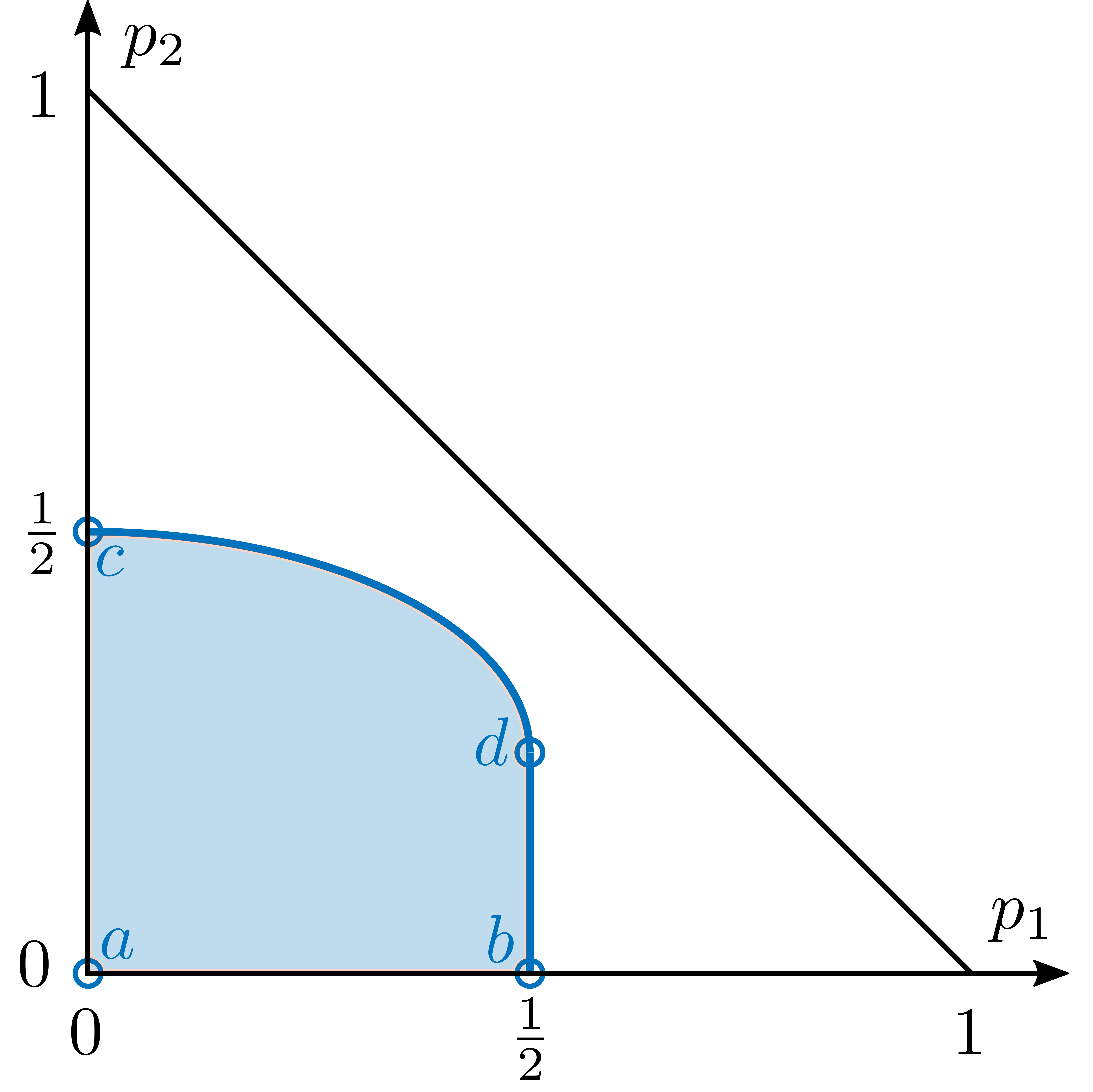}
	\caption{Two-dimensional representation (white triangle) of quantum states of the form \eqref{eq:mixture_2_photons}. The blue region included in the triangle corresponds to the Wigner-positive states. The points $a$, $b$, $c$, and $d$ are associated with the Wigner functions $W_a$, $W_b$, $W_c$, and $W_d$, while the points in the segment of an ellipse connecting $c$ and $d$ have a Wigner function $V_t$ where $t$ takes any value between $0$ and $1$. Taken from \cite{VanHerstraeten2020}. }
	\label{fig:set_wig_pos_2D}
\end{figure}

In this Section, we make a first step towards solving conjecture \eqref{eq:conjW0majW+} by considering a particular subset of Wigner-positive states, namely phase-invariant  states that are restricted to two photons at most,
\begin{equation}
	\rho = 
	\left(
	1-p_1-p_2
	\right)
	\ket{0}\bra{0}
	+p_1\ket{1}\bra{1}
	+p_2\ket{2}\bra{2},
	\label{eq:mixture_2_photons}
\end{equation}
where $p_1,p_2 \geq 0$ and $p_1+p_2 \leq 1$. These states form a convex set given by the area whose outer boundaries are the $p_1$-axis, the $p_2$-axis and the line satisfying $p_1+p_2=1$ as pictured on Fig.~\ref{fig:set_wig_pos_2D}.
We will prove conjecture \eqref{eq:conjW0majW+} for the set of Wigner-positive states of the form \eqref{eq:mixture_2_photons}, denoted as $\mathcal{W}_+^\mathrm{restr}$, which has previously been studied in \cite{Brocker1995,VanHerstraeten2020}. It is obvious to see that $\mathcal{W}_+^\mathrm{restr}$ forms a convex set as well since any convex mixture of Wigner-positive states is Wigner-positive, but the boundary of this set is nonetheless non trivial, see Fig. \ref{fig:set_wig_pos_2D}. At the same time, this set is simple enough to enable a fully analytical proof of the conjecture \eqref{eq:conjW0majW+}.

As shown in \cite{VanHerstraeten2020}, the boundary of the set $\mathcal{W}_+^\mathrm{restr}$ comprises the extremal states $\rho_a$, $\rho_b$, $\rho_c$ and $\rho_d$, represented by the corresponding letters in Fig. \ref{fig:set_wig_pos_2D}, as well as the segment of an ellipse connecting $\rho_c$ to $\rho_d$.
Thus, any state in  $\mathcal{W}_+^\mathrm{restr}$ can be written as a convex mixture of these extremal states. Note that $\rho_a=\ket{0}\bra{0}$, which lies at the origin in Fig. \ref{fig:set_wig_pos_2D}, is simply the vacuum state which will be proven to majorize every other state.
The expressions of the Wigner functions of the first four extremal states (as a function of the parameter $r$) read as follows \cite{VanHerstraeten2020}:
\begin{equation}
	\begin{split}
		W_a(r) &= W_0(r) = \dfrac{1}{\pi}\exp\left(-r^2\right),
		\\
		W_b(r) &= \dfrac{1}{\pi}\exp\left(-r^2\right)r^2,
		\\
		W_c(r) &= \dfrac{1}{\pi}\exp\left(-r^2\right)\left(r^2-1\right)^2,
		\\
		W_d(r) &= \dfrac{1}{\pi}\exp\left(-r^2\right)\dfrac{1}{2}r^4.
	\end{split}
\label{eq:wigner_func_extremal_wigpos}
\end{equation}
In addition to these, there is a continuum of extremal states located on the segment of an ellipse connecting point $c$ to point $d$ in Fig. \ref{fig:set_wig_pos_2D}. Using a parameter $t \in [0,1]$, the Wigner function of these states can be parametrized as follows \cite{VanHerstraeten2020}:
\begin{equation}
	V_t(r)=
	\dfrac{t+1}{2 \pi}\exp\left(-r^2\right)
	\left(r^2-1+\sqrt{\dfrac{1-t}{1+t}}\right)^2.
\end{equation}
Note that for $t=0$, $V_t$ coincides with $W_d$, while for $t=1$, it coincides with $W_c$.

We are now going to prove that Eq. \eqref{eq:conjW0majW+} holds for all states contained in the convex set $\mathcal{W}_+^\mathrm{restr}$. To do so, it is sufficient to prove that the Wigner functions of all the extremal states are majorized by the Wigner function of the vacuum $W_0$. As a consequence of Eq. \eqref{eq:majorization_convex_mixture}, this will indeed automatically imply that the same majorization relation holds for all convex mixtures of extremal states, hence for all states in $\mathcal{W}_+^\mathrm{restr}$.
In order to prove our result, we begin by showing that a majorization relation on radial functions in $\mathbb{R}^n$ (here, we only need $n=2$) is equivalent to a majorization relation on specific functions defined on the non-negative real line. This is the content the following lemma, which we prove in Appendix \ref{app:lem:major-1dim}.
\begin{lemma} \label{lem:major-1dim}
	If $f$ and $g$ are two $n$-dimensional radial distributions defined on $\mathbb{R}^n$ such that $f(\mathbf{r})=\fr\left(\Vert\mathbf{r}\Vert\right)$ and $g(\mathbf{r})=\gr\left(\Vert\mathbf{r}\Vert\right)$ with $\fr$ and $\gr$ defined on $\mathbb{R}_+$, then $f\succ g$ is equivalent to $\tilde{f}\succ \tilde{g}$, where $\tilde{f}$ and $\tilde{g}$ are one-dimensional distributions defined on $\mathbb{R}_+$ as $\tilde{f}(x) = \fr\left(\sqrt[n]{x}\right)$ and $\tilde{g}(x) = \gr\left(\sqrt[n]{x}\right)$.
\end{lemma}
\noindent Lemma \ref{lem:major-1dim} implies that a majorization relation between any two Wigner functions picked from $W_0$, $W_b$, $W_c$, $W_d$ and $V_t$ is equivalent to a majorization relation between the corresponding one-dimensional functions picked from $f_0$, $f_b$, $f_c$, $f_d$, and $g_t$, which are defined on $\mathbb{R}_+$ as
\begin{equation}
	\begin{split}
		f_0(x) &= \exp\left(-x\right),
		\\
		f_b(x) &= \exp\left(-x\right)x,
		\\
		f_c(x) &= \exp\left(-x\right)\left(x-1\right)^2,
		\\
		f_d(x) &= \exp\left(-x\right)\dfrac{1}{2}x^2,
	\end{split}
	\label{eq:expression_f0_fb_fc_fd}
\end{equation}
and
\begin{equation}
	g_t(x)
	=
	\exp\left(-x\right)
	\dfrac{1}{2}(t+1)
	\left(x-1+\sqrt{\dfrac{1-t}{1+t}}\right)^2.
	\label{eq:expression_ft}
\end{equation}

Thus, we need to prove now that $f_0$ majorizes $f_b$, $f_c$, $f_d$, and $g_t$. Our proof relies on the following lemma, which we prove in Appendix \ref{app:lem:mixture_of_distributions} for completeness, as we could not find it in the literature. 
\begin{lemma} \label{lem:mixture_of_distributions}
	Consider two probability distributions $f$ and $g$ defined on the same domain $\mathcal{A}$. If there exists a collection of level-equivalent distributions $f^{(\alpha)}$ on $\mathcal{A}$ depending on the parameter $\alpha \in \Omega$ with $f^{(\alpha)}\equiv f$ for all $\alpha$ such that
	\begin{equation}
		g\left(\mathbf{r}\right) = 
		\int_\Omega
		f^{(\alpha)}\left(\mathbf{r}\right)\, \mathrm{d}k(\alpha), \quad \forall \, \mathbf{r} \in \mathcal{A},
	\end{equation}
	where $k:\Omega\mapsto\mathbb{R}_+$ is a probability measure on $\Omega$, 
	then $f\succ g$.
\end{lemma}
\noindent Lemma \ref{lem:mixture_of_distributions} enables us to prove that $f\succ g$ provided that we can build $g$ as some convex mixture of distributions that are level-equivalent to $f$. Note that it is very similar in its spirit to the characterization of discrete majorization in terms of convex mixtures of permutations \cite{Marshall2011}. However, while the latter gives a necessary and sufficient condition, Lemma \ref{lem:mixture_of_distributions} only provides a sufficient condition for majorization. Although it is unknown, to our knowledge,  whether it could be promoted to an equivalence (similarly to Proposition \ref{prop:semi-doubly-stochastic}), a sufficient condition is all we need to prove the results of our paper.


\subsection*{Case of $f_b$ and $f_d$}

Let us first prove that $f_0$ majorizes $f_b$ and $f_d$. In order to make use of Lemma \ref{lem:mixture_of_distributions}, we are going to build an appropriate collection of  level-equivalent functions to $f_0$. One simple way to generate level-equivalent functions is simply by shifting the original function to the right. Starting from $f_0$, we define the functions $f_0^{(\alpha)}$ labelled by the non-negative shift parameter $\alpha$ as
\begin{equation}
	f_0^{(\alpha)}(x) = \exp(-x+\alpha) \, \Theta(x-\alpha),
\end{equation}
where $\Theta(z)$ represents the Heaviside step function.
We obviously have that $f_0^{(\alpha)} \equiv f_0$ for all $\alpha \geq 0$.
Now, define the probability densities $k_b(\alpha)=\exp(-\alpha)$ and $k_d(\alpha)=\alpha\, \exp(-\alpha)$, with $\alpha\in\mathbb{R}_+$. It is trivial to verify that $k_{b(d)}(\alpha)\geq 0$ for all $\alpha \in \mathbb{R}_+$ and $\int k_{b(d)}(\alpha)\, \mathrm{d}\alpha=1$. Furthermore, it can easily be shown that
\begin{equation}
		f_{b(d)}(x) = \int\limits_{0}^{+\infty}k_{b(d)}(\alpha) \, f_0^{(\alpha)}(x) \, \mathrm{d}\alpha.
\end{equation}
Lemma \ref{lem:mixture_of_distributions} then directly implies that $f_0\succ f_b$ and $f_0\succ f_d$, where we have used the probability measure $k_{b(d)}$ over $\Omega=\mathbb{R}_+$.

\subsection*{Case of $f_c$ and $g_t$}

\begin{figure}
	\includegraphics[width=\linewidth]{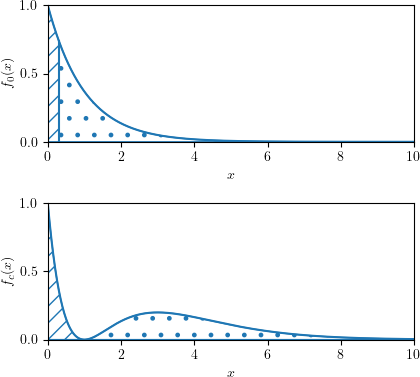}
	\caption{Illustration of the ``split'' of $f_0$ (top) and $f_c$ (bottom) into two parts. The hashed parts are $f_0^-$ and $f_c^-$, while the dotted parts are $f_0^+$ and $f_c^+$. The two hashed areas are equal, and the two dotted areas are equal. This allows us to treat  $f^-_0\succ f^-_c$ and $f^+_0\succ f^+_c$ separately, in order to conclude finally that $f_0 \succ f_c$.}
	\label{fig:f0_and_fc}
\end{figure}

The same method is not directly applicable to prove that $f_0$ majorizes $f_c$ and  $g_t$ because the latter functions are non-zero at the origin. The trick, however, is to exploit the fact that  $f_c$ looks like a rescaled version of $f_d$ in the domain $[1, \infty)$. We can then ``split" $f_c$ into two parts and prove the majorization relation separately for each part. This is possible as a consequence of the following lemma, which we prove in Appendix \ref{app:lem:major_tensor}.
\begin{lemma} \label{lem:major_tensor}
	Consider four functions $f_1$, $f_2$, $g_1$, and $g_2$ defined on the same domain $\mathcal{A}$ and such that $f_1$ and $f_2$ do not both take non-zero values in the same element of $\mathcal{A}$, and similarly $g_1$ and $g_2$ do not both take non-zero values in the same element of $\mathcal{A}$. If the functions satisfy $f_1\succ g_1$ and $f_2\succ g_2$, then $(f_1+f_2)\succ (g_1+g_2)$.
\end{lemma}
\noindent In light of Lemma \ref{lem:major_tensor}, define the two functions $f_c^-$ and $f_c^+$ on $\mathbb{R}_+$ as
\begin{equation}
	\begin{split}
		f_c^-(x) = 
		\begin{cases}
			f_c(x),
			\quad
			&\text{for } 0\leq x\leq 1,
			\\
			0,\quad
			&\text{else},
		\end{cases}
	\end{split}
\end{equation}
and
\begin{equation}
	\begin{split}
		f_c^+(x) = 
		\begin{cases}
			0,
			\quad
			&\text{for } 0\leq x\leq 1,
			\\
			f_c(x),
			\quad
			&\text{else}.
		\end{cases}
	\end{split}
\end{equation}
Obviously, we have $f^-_c+f^+_c=f_c$. In order to prove that $f_0\succ f_c$ by using Lemma \ref{lem:major_tensor}, we also need to ``split" $f_0$ into two parts $f_0^-$ and $f_0^+$ such that $f^-_0+f^+_0=f_0$. Moreover, in order to be able to apply majorization on each part, $f_0^-$ must have the same normalization as $f_c^-$, and similarly for $f_0^+$ and $f_c^+$. Define $x^* = 1 - \ln 2$, and note that
\begin{equation}
	\begin{split}
	\int\limits_{0}^{1}f_c(x)\mathrm{d}x=
	&\int\limits_{0}^{x^*}f_0(x)\mathrm{d}x,
	\\
	\int\limits_{1}^{\infty}f_c(x)\mathrm{d}x=
	&\int\limits_{x^*}^{\infty}f_0(x)\mathrm{d}x.
	\end{split}
	\label{eq:norm_f0_fc}
\end{equation}
With this in mind, the functions $f_0^-$ and $f_0^+$ on $\mathbb{R}_+$ are
\begin{equation}
	f_0^-(x) =
	\begin{cases}
		f_0(x), \quad
		&\text{for }0\leq x\leq x^*,
		\\
		0,
		&\text{else},
	\end{cases}
\end{equation}
and
\begin{equation}
	f_0^+(x) =
	\begin{cases}
		0,\quad
		&\text{for }0\leq x\leq x^*,
		\\
		f_0(x),\quad
		&\text{else}.
	\end{cases}
\end{equation}
It follows from \eqref{eq:norm_f0_fc} that $f^-_c$ and $f_0^-$ have the same normalization (and similarly for $f^+_c$ and $f_0^+$). The distributions $f^-_0$, $f^+_0$, $f^-_c$, and $f^+_c$ are represented in Fig. \ref{fig:f0_and_fc}.

The last step now is to prove that $f^-_0\succ f^-_c$ as well as $f^+_0\succ f^+_c$. Starting  with the latter relation, we define the two functions $\tilde{f}^+_0$ and $\tilde{f}^+_c$ on $\mathbb{R}_+$ by respectively shifting to the left $f^+_0$ by an amount $x^*$ and $f^+_c$ by $1$, namely
\begin{equation}
	\tilde{f}^+_0(x)
	=
	f^+_0\left(x+1-\ln 2\right)
	=
	2
	\exp(-1)
	\exp(-x),
\end{equation}
and
\begin{equation}
	\tilde{f}^+_c(x)
	=
	f^+_c\left(x+1\right)
	=
	\exp(-1)
	\exp(-x)
	x^2.
\end{equation}
Note that $\tilde{f}^+_0(x)$ and $\tilde{f}^+_c(x)$ are proportional to $f_0(x)$ and $f_d(x)$, respectively, with the same proportionality factor of $2\exp(-1)$. Since we have already shown that $f_0\succ f_d$, it follows that $\tilde{f}^+_0\succ \tilde{f}^+_c$, which is equivalent to $f_0^+\succ f_c^+$ since $\tilde{f}_0^+\equiv f_0^+$ and $\tilde{f}_c^+\equiv f_c^+$.

In order to prove that  $f^-_0\succ f^-_c$, we note that $f^-_0$ and $f^-_c$ are both monotonically decreasing functions, so they coincide with their decreasing rearrangements, namely  $f^-_0=(f^-_0)^{\downarrow}$ and $f^-_c=(f^-_c)^{\downarrow}$. Therefore, their cumulative integrals are simply given by
\begin{equation}
	S_{f^-_0}(s) = \int\limits_{0}^{s}f^-_0(x)\,  \mathrm{d}x
	\quad
	\text{and}
	\quad
	S_{f^-_c}(s) = \int\limits_{0}^{s}f^-_c(x)\,  \mathrm{d}x.
\end{equation}
In order to prove the majorization relation, we will now show that $S_{f^-_0}(s) \geq S_{f^-_c}(s)$ for all $s \in \mathbb{R}_+$. Since $f^-_0$ and $f^-_c$ are both monotonically decreasing functions, since $f^-_0(x) = 0$ for all $x>x^*$, and since $x^*<1$, it is sufficient to show that $f^-_0(x) \geq f^-_c(x)$ for all $x \in [0,x^*]$. Indeed, the ratio $f_c(x)/f_0(x)=(x-1)^2 \le 1$ for $x \in [0,x^*]$.
Using Definition \ref{def:def_major} of a majorization relation, we get $f^-_0 \succ f^-_c$. From Lemma \ref{lem:major_tensor}, we conclude that $f_0 \succ f_c$.

Finally, the same ``splitting" technique can be used to prove that $f_0 \succ g_t$ for all values of $t \in [0,1]$, which of course includes $f_0 \succ f_d$ and $f_0 \succ f_c$ as limiting cases for $t=0$ and $1$, respectively. We point the interested reader to Appendix \ref{apd:parabola} for such a proof. 
In summary, we have thus shown that all functions (i.e., $f_b$, $f_c$, $f_d$ and  $g_t$ for all $t \in [0,1]$) are majorized by $f_0$.
Using Lemma \ref{lem:major-1dim}, this translates into the fact that the Wigner functions of all extremal states (i.e., $W_b$, $W_c$, $W_d$, and $V_t$ for all $t \in [0,1]$) are majorized by $W_0$. Hence, any convex mixture of these extremal Wigner functions is also majorized by $W_0$ as a consequence of Eq. \eqref{eq:majorization_convex_mixture}.
This concludes the proof of Eq. \eqref{eq:conjW0majW+} for all Wigner-positive states in $\mathcal{W}_+^\mathrm{restr}$.

\section{Discussion and conclusion}
\label{sec:conclusion}

In the present work, we have shown that continuous majorization theory proves to be an elegant and powerful tool for exploring the information-theoretic properties of Wigner functions representing quantum states in phase space. While it only applies to states admitting a positive Wigner function, continuous majorization should nevertheless pave the way to the proof of various entropic inequalities of interest in quantum physics. This is so because, as explained in Sec. \ref{sec:cont_major}, a continuous majorization relation implies an infinite set of entropic inequalities for the Wigner functions involved. 

For instance, an interesting application concerns the entropic uncertainty relation due to Bia{\l}ynicki-Birula and Mycielski \cite{Bialynicki1975}, which can be viewed as the entropic counterpart of the Heisenberg uncertainty relation $\sigma_x \sigma_p \geq 1/2$, where $\sigma_x$ and $\sigma_p$ denote the standard deviation of the $x$-distribution $\rho_x$ and $p$-distribution $\rho_p$ of a quantum state. The entropic uncertainty relation reads $h\left(\rho_x\right)+h\left(\rho_p\right)\geq\ln\pi+1$, where the right-hand side is simply the Shannon (differential) entropy of the Wigner function of the vacuum, which is also equal to the sum of the entropies of its marginals. This inequality is, however, not fully satisfying as it is not saturated for all Gaussian pure states \cite{Hertz2017}. For Wigner-positive states, the bound on the Wigner entropy  $h(W)\geq h(W_0) = \ln\pi+1$ has been conjectured -- and proven in some cases -- in \cite{VanHerstraeten2020}. Since $h(W)=h\left(\rho_x\right)+h\left(\rho_p\right)-I(W)$, where $I(W) \ge 0$ is the mutual information between $x$ and $p$, this bound on the Wigner entropy yields a stronger entropic uncertainty relation in the presence of $x$-$p$ correlations. Remarkably, our conjecture  \eqref{eq:conjW0majW+} that $W_0 \succ W$ for all Wigner-positive states in $\mathcal{W}_+$ then immediately implies this strong bound on the Wigner entropy \footnote{The bound $h(W)\ge h(W_0)$ was proven in \cite{VanHerstraeten2020} for a subset of Wigner-positive states that are called passive states, which are especially relevant in quantum thermodynamics. In the restricted case of states with two photons at most, this corresponds to some triangle included in the blue region of Fig.  \ref{fig:set_wig_pos_2D}, see \cite{VanHerstraeten2020} for more details. In contrast, we have proven here that $W_0 \succ W$ which immediately implies  $h(W)\ge h(W_0)$ for all states in the blue region.}. It actually also implies similar bounds on all R\'enyi entropies $h_{\alpha}(W)$ as well as on all concave functionals $\phi(W)$ as defined in Eq. \eqref{eq:phi_int_of_phi}. This clearly illustrates how majorization theory can serve as an efficient tool towards the strengthening of entropic uncertainty relations.

\begin{figure}[t]	\includegraphics[width=\linewidth]{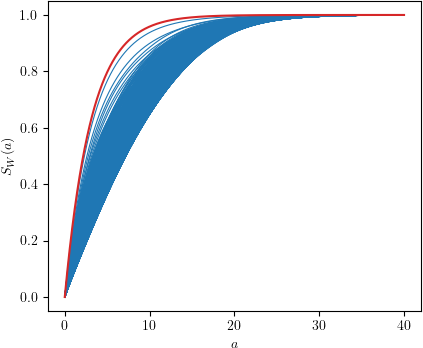}
	\caption{Cumulative integral $S_W$ as a function of the area parameter $a=\pi s^2$ for positive Wigner functions $W$. The red curve represents the cumulative integral for $W_0$, while each blue curve represents the cumulative integral for a random Wigner-positive state.
	In total, 1000 random instances of Wigner-positive states are uniformly sampled over the set of mixed states containing up to 2 photons (more details on the numerics are given in Appendix \ref{apd:numerics}).
	The red curve lies above all blue curves, confirming conjecture \eqref{eq:conjW0majW+}.}
	\label{fig:numCumSum}
\end{figure}

\begin{figure}[t]
	\includegraphics[width=\linewidth]{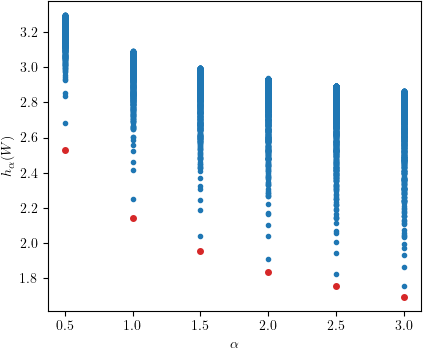}
	\caption{R\'enyi entropy $h_\alpha(W)$ for positive Wigner functions $W$ as a function of the parameter $\alpha$. The red points represent the R\'enyi entropy of $W_0$, while the blue points represent the R\'enyi entropy of random positive Wigner functions. 
	In total, 1000 random instances of positive Wigner functions are uniformly sampled over the set of mixed states containing up to 2 photons (more details on the numerics are given in Appendix \ref{apd:numerics}).
	Since the R\'enyi entropy $h_\alpha$ is Schur-concave for all $\alpha \in \mathbb{R_+}$, Eq. \eqref{eq:conjW0majW+} implies that $h_\alpha(W) \ge h_\alpha(W_0)$ for all positive Wigner functions $W$, as can be verified in this figure.}
	\label{fig:numRenyi}
\end{figure}

In this paper, we have proven the fundamental majorization relation $W_0 \succ W$ restricted to states in the set $\mathcal{W}_+^\mathrm{restr}$ of phase-invariant states with two photons at most. We believe the same techniques laid out here could be exploited to prove our majorization relation for all states in $\mathcal{W}_+$. We choose, however, to leave this investigation for future work but we provide numerical evidences supporting our conjecture. In Figure \ref{fig:numCumSum}, we plot the cumulative integral of the Wigner function $W_0$ and compare it with the cumulative integrals of the Wigner functions $W$ of some randomly chosen (hence, non-Gaussian) Wigner-positive states. We see that $S_{W_0}(s) \ge S_{W}(s)$ for all $s\ge 0$, which confirms that $W_0 \succ W$  in view of Definition \ref{def:def_major}. In Figure \ref{fig:numRenyi}, we show that the R\'enyi entropy $h_{\alpha}$ of the Wigner function $W_0$ is smaller, for several values of the parameter $\alpha$, than its value for the same Wigner functions $W$. This is again consistent with $W_0 \succ W$ in view of Definition \ref{def:Shur_concave} and the fact that  $h_{\alpha}$ is Schur-concave.	
The interested reader can find the details on our numerical method in Appendix \ref{apd:numerics}.

Interestingly, our conjecture \eqref{eq:conjW0majW+} bears some resemblance with the so-called generalized Wehrl's conjecture \cite{wehrl1978general}. It has indeed been proven by Lieb and Solovej \cite{lieb2002proof,Lieb2014} that any concave function of the Husimi Q-function of a state is lower bounded by the same function applied to the Husimi Q-function of the vacuum state (or any coherent state). This is actually equivalent to stating that the Husimi function of the vacuum majorizes the Husimi function of any other state. Intriguingly, this is not proved in \cite{Lieb2014} with continuous majorization, but using discrete majorization for finite-dimensional spin-coherent states followed by some limiting argument. Following the same line of thought as in \cite{VanHerstraeten2020}, our conjecture \eqref{eq:conjW0majW+} actually implies the generalized Wehrl's conjecture (since any Husimi function is also the Wigner function of another physical state) but the converse is not true as it is easy to produce positive Wigner functions that do not coincide with the Husimi function of a physical state.

Beyond proving our majorization relation \eqref{eq:conjW0majW+}, the next step would of course be to extend it to all Wigner-positive states in arbitrary dimensions. 
We conjecture that any $N$-mode Wigner-positive state has a Wigner function that is majorized by that of any $N$-mode pure Gaussian state.
A more challenging direction of research would then be to account for partly negative Wigner functions. One way of doing so could be to apply a majorization relation to a carefully chosen non-negative distribution that characterizes any Wigner function. Another way could be to extend the notion of majorization to partly negative functions defined on $\mathbb{R}^n$ (this is known to be possible for functions defined on a finite domain \cite{Marshall2011}), which would imply proper inequalities involving the entropies of the marginal of any Wigner function. 
However, we do not expect a straightforward extension of our majorization conjecture to exist for Wigner-negative states. Indeed, consider the negative volume of a Wigner function defined as $\mathrm{Vol}_-(W)=-\iint \left[W(x,p)\right]_-\mathrm{d}x\, \mathrm{d}p$. It is easily seen that the function $\varphi(x)=-\left[x\right]_-=-\min(x, 0)$ is convex, so that $\mathrm{Vol}_-(\cdot)$ is Schur-convex. As a consequence, we understand that any Wigner function with non-zero negative volume cannot be majorized by a Wigner function with zero negative volume, such as the one of a Gaussian pure state.
Overall, we anticipate that the application of continuous majorization theory to Wigner functions  will prove very fruitful for elucidating the properties of quantum states in phase space.

\section*{Acknowledgements}
We thank Anaelle Hertz for helpful discussions on this subject. Z.V.H. acknowledges a fellowship from the FRIA foundation (F.R.S.-FNRS). M.G.J. acknowledges a fellowship from the Wiener-Anspach Foundation. N.J.C. acknowledges support by the F.R.S.-FNRS under Project No. T.0224.18 and by the EC under project ShoQC within ERA-NET Cofund in Quantum Technologies (QuantERA) program.

\appendix

\section{Proofs of some majorization lemmas} \label{apd:major_lemmas}

\subsection{Proof of Lemma \ref{lem:major-1dim}} \label{app:lem:major-1dim}

Using polar coordinates, we have that $f\succ g$ is equivalent to the following condition:
\begin{equation}
	\int_{0}^{\infty} \left[\fr(r)-t\right]_+ 
	r^{n-1}
	\, \mathrm{d}r
	\geq
	\int_{0}^{\infty} \left[\gr(r)-t\right]_+
	r^{n-1}
	\, \mathrm{d}r .
\end{equation}
Introducing the change of variables $x = r^n$, we rewrite the above inequality as
\begin{equation}
	\int_{0}^{\infty} \left[\fr\left(\sqrt[n]{x}\right)-t\right]_+ \mathrm{d}x
	\geq
	\int_{0}^{\infty} \left[\gr\left(\sqrt[n]{x}\right)-t\right]_+ \mathrm{d}x.
\end{equation}
Defining $\tilde{f}(x) = \fr\left(\sqrt[n]{x}\right)$ and $\tilde{g}(x) = \gr\left(\sqrt[n]{x}\right)$, it follows that $\tilde{f}\succ\tilde{g}$ is equivalent to $f\succ g$.

\subsection{Proof of Lemma \ref{lem:mixture_of_distributions}} \label{app:lem:mixture_of_distributions}

Since the function $\gamma_t(z)=\left[z-t\right]_+$ is convex, we can exploit Jensen's inequality to get
\begin{equation}
	\begin{split}
		\gamma_t\left(g\left(\mathbf{r}\right)\right)
		&=
		\gamma_t\left(
		\int_{\Omega}
        \,f^{(\alpha)}\left(\mathbf{r}\right)\mathrm{d}k(\alpha)
		\right)
		\\
		&\leq
		\int_{\Omega}
		\, 
		\gamma_t\left(
		f^{(\alpha)}\left(\mathbf{r}\right)
		\right)
		\mathrm{d}k(\alpha).
	\end{split}
\end{equation}
Integrating both terms over the domain $\mathcal{A}$ then leads to
\begin{equation}
	\begin{split}
		\int_{\mathcal{A}}
		\gamma_t\left(g\left(\mathbf{r}\right)\right)
		\mathrm{d}\mathbf{r}
		&\leq
		\int_{\Omega} 
		\int_{\mathcal{A}}
		\gamma_t\left(
		f^{(\alpha)}\left(\mathbf{r}\right)
		\right)
				\mathrm{d}\mathbf{r}  \,
		\mathrm{d}k(\alpha).
	\end{split}
\end{equation}
Since $f$ and $f^{(\alpha)}$ are level-equivalent, we have that 
\begin{equation} \label{eq:lem:app}
	\int\gamma_t
	\left(
	f^{(\alpha)}
	\left(\mathbf{r}\right)
	\right)
	\mathrm{d}\mathbf{r}
	=
	\int\gamma_t
	\left(
	f
	\left(\mathbf{r}\right)
	\right)
	\mathrm{d}\mathbf{r}.
\end{equation}
The integral over $k(\alpha)$ reduces to $1$ since it is a probability measure. Writing $\gamma_t$ explicitly, we end up with
\begin{equation}
	\int 
	\left[
	f(\mathbf{r})-t
	\right]_+
	\mathrm{d}\mathbf{r}
	\geq
	\int 
	\left[
	g(\mathbf{r})-t
	\right]_+
	\mathrm{d}\mathbf{r},
\end{equation}
which proves the Lemma.

\subsection{Proof of Lemma \ref{lem:major_tensor}} \label{app:lem:major_tensor}

From the fact that for each $\mathbf{r}$, either $f_1(\mathbf{r})$ or $f_2(\mathbf{r})$ is equal to zero, it follows that :
\begin{equation}
	\left[
	f_1(\mathbf{r})-t
	\right]_+
	+
	\left[
	f_2(\mathbf{r})-t
	\right]_+
	=
	\left[
	f_1(\mathbf{r})+f_2(\mathbf{r})-t
	\right]_+,
\end{equation}
since $t\geq 0$. The same applies to $g_1$ and $g_2$. Let $f = f_1+f_2$ and $g=g_1+g_2$. We have:
\begin{equation}
	\begin{split}
		&\int\left[
		f(\mathbf{r})-t
		\right]_+
		\mathrm{d}\mathbf{r}
            \\
		=&
		\int\left[
		f_1(\mathbf{r})+f_2(\mathbf{r})-t
		\right]_+
		\mathrm{d}\mathbf{r}
		\\
		=&
		\int\left[
		f_1(\mathbf{r})-t
		\right]_+
		\mathrm{d}\mathbf{r}
		+
		\int\left[
		f_2(\mathbf{r})-t
		\right]_+
		\mathrm{d}\mathbf{r}
		\\
		\geq&
		\int\left[
		g_1(\mathbf{r})-t
		\right]_+
		\mathrm{d}\mathbf{r}
		+
		\int\left[
		g_2(\mathbf{r})-t
		\right]_+
		\mathrm{d}\mathbf{r}
		\\
		=&
		\int\left[
		g_1(\mathbf{r})+g_2(\mathbf{r})-t
		\right]_+
		\mathrm{d}\mathbf{r}
		\\
		=&
		\int\left[
		g(\mathbf{r})-t
		\right]_+
		\mathrm{d}\mathbf{r},
	\end{split}
\end{equation}
where the inequality follows from $f_1\succ g_1$ and $f_2\succ g_2$.

\section{Proof of the majorization relation for the states located on the ellipse}
\label{apd:parabola}

In this Appendix, we prove that the Wigner functions of the extremal Wigner-positive states located on the ellipse represented in Fig. \ref{fig:set_wig_pos_2D} are majorized by the Wigner function of the vacuum state, by showing that \ $f_0\succ g_t$ for all $t$ such that $0\leq t\leq 1$. Note that the proof is very similar to the proof of $f_0\succ f_c$. The function $g_t(x)$ defined in Eq. \eqref{eq:expression_ft} has one zero at $x = a_t$, where
\begin{equation}
	a_t = 1-\sqrt{\dfrac{1-t}{1+t}}.
\end{equation}
We ``split" $g_t$ in two different functions $g^-_t$ and $g^+_t$ from either sides of $a_t$:
\begin{equation}
	\begin{split}
		g_t^-(x) = 
		\begin{cases}
			g_t(x),
			\quad
			&\text{for } 0\leq x\leq a_t,
			\\
			0,\quad
			&\text{else},
		\end{cases}
	\end{split}
\end{equation}
\begin{equation}
	\begin{split}
		g_t^+(x) = 
		\begin{cases}
			0,
			\quad
			&\text{for } 0\leq x\leq a_t,
			\\
			g_t(x),
			\quad
			&\text{else}.
		\end{cases}
	\end{split}
\end{equation}
If we shift $g^+_t$ from $a_t$ towards the origin, we have a distribution that is proportional to $\exp(-x)x^2/2$, which we know is majorized by $f_0(x)=\exp(-x)$ from our previous result.
The idea now is to split $f_0$ in two different functions with the same normalization as $g^-_t$ and $g^+_t$. Define $b_t$ as
\begin{equation}
	b_t = 1-\ln(1+t)-\sqrt{\dfrac{1-t}{1+t}}.
\end{equation}
It satisfies
\begin{equation}
	\begin{split}
		\int\limits_{0}^{a_t}g_t(x)\mathrm{d}x=&
		\int\limits_{0}^{b_t}f_0(x)\mathrm{d}x,
		\\
		\int\limits_{a_t}^{\infty}g_t(x)\mathrm{d}x=&
		\int\limits_{b_t}^{\infty}f_0(x)\mathrm{d}x.
	\end{split}
	\label{eq:bt_def_integral}
\end{equation}
We now ``split" $f_0$ from either side of $x=b_t$, and define $f_0^-$ and $f_0^+$ as:
\begin{equation}
	f_0^-(x) =
	\begin{cases}
		f_0(x),\quad
		&\text{for }0\leq x\leq b_t,
		\\
		0,
		&\text{else},
	\end{cases}
\end{equation}
\begin{equation}
	f_0^+(x) =
	\begin{cases}
		0,\quad
		&\text{for }0\leq x\leq b_t,
		\\
		f_0(x),\quad
		&\text{else}.
	\end{cases}
\end{equation}
We are now going to prove that $f^-_0\succ g^-_t$ and $f^+_0\succ g^+_t$.
Let us shift $g^+_t$ and $f^+_0$ towards the origin respectively from $a_t$ and $b_t$. We end up with the functions:
\begin{equation}
	\begin{aligned}
	\tilde{g}^+_t(x)
	& =
	g^+_t\left(x+a_t\right) \\
	& =
	(t+1)
	\exp\left(\sqrt{\dfrac{1-t}{1+t}}-1\right)
	\exp(-x)
	\dfrac{1}{2}
	x^2,
	\end{aligned}
\end{equation}
and
\begin{equation}
	\begin{aligned}
	\tilde{f}^+_0(x)
	& =
	f^+_0\left(x+b_t\right) \\
	& =
	(t+1)
	\exp\left(\sqrt{\dfrac{1-t}{1+t}}-1\right)
	\exp(-x).
	\end{aligned}
\end{equation}
$\tilde{f}^+_0(x)$ and $\tilde{g}^+_t(x)$ are respectively proportional to $\exp(-x)$ and $\exp(-x)x^2/2$, with the same proportionality factor. Since we have already shown that $\exp(-x)\succ\exp(-x)x^2/2$, it follows that $\tilde{f}^+_0\succ \tilde{g}^+_t$.
Since $\tilde{f}_0^+$ and $\tilde{g}_t^+$ are level-equivalent to respectively $f_0^+$ and $g_t^+$, we then have $f_0^+\succ g_t^+$.

We now turn to $f^-_0\succ g^-_t$. They are both monotonically decreasing, making them decreasing rearrangements. Therefore their cumulative integrals are given by:
\begin{equation}
	S_{f^-_0}(s) = \int\limits_{0}^{s}f^-_0(x)\mathrm{d}x
	\quad
	\text{and}
	\quad
	S_{g^-_t}(s) = \int\limits_{0}^{s}g^-_t(x)\mathrm{d}x.
\end{equation}
We will now show that $S_{f^-_0}(s) \geq S_{g^-_t}(s)$ for all $s \in \mathbb{R}_+$. Since $f^-_0$ and $g^-_t$ are both monotonically decreasing and $f^-_0(x) = 0$ for all $x>b_t$ coupled with the fact that $b_t<a_t$, it is sufficient to show that $f^-_0(x) \geq g^-_t(x)$ for all $x \in [0,b_t]$. To prove this, note that the ratio $g_t(x)/f_0(x)$ is less than $1$ for $x \in [0,b_t]$:
\begin{equation}
	\dfrac{g_t(x)}{f_0(x)}
	=
	\dfrac{t+1}{2}\left(x-1+\sqrt{\dfrac{1-t}{1+t}}\right)^2\leq 1,
\end{equation}
which follows from the fact that $x\leq b_t \leq 1$ and $0\leq t\leq 1$. Using Definition \ref{def:def_major} of a majorization relation, we conclude that $f^-_0 \succ g^-_t$. From Lemma \ref{lem:major_tensor}, we finally end up with $f_0 \succ g_t$.

\section{Numerical exploration of the majorization conjecture}
\label{apd:numerics}

In this Appendix, we present our numerical results supporting conjecture \eqref{eq:conjW0majW+}, namely that the Wigner function of the vacuum majorizes any non-negative Wigner function.
We begin by describing how we uniformly sample the set of Wigner-positive states in a truncated Fock basis.
We then give some details on how we numerically check a majorization relation between two Wigner functions.
Finally, we exhibit additional plots illustrating the validity of the conjecture.

\subsection{Truncated Fock basis}
The basis of Fock states can be used to decompose any quantum state described by a density operator $\hat{\rho}$:
\begin{equation}
	\hat{\rho} = 
	\sum\limits_{k=0}^{\infty}\sum\limits_{l=0}^{\infty}
	\rho_{kl}\ket{k}\bra{l},
\end{equation}
where $\{ \ket{k}\}_{k \in \mathbb{N}_0}$ represents the Fock basis.
Since the matrix $\rho$ represents a physical quantum state, it is a trace-one, Hermitian, and positive semi-definite matrix. 
One can compute the Weyl transform of every operator $|k\rangle\langle l|$, which we denote as $W_{kl}$, as follows:
\begin{equation}
	W_{kl}(x,p) 
	= 
	\dfrac{1}{\pi}
	\int
	\exp\left(2ipy\right)
	\psi_{k}^{\ast}\left(x+y\right)
	\psi_{l}\left(x-y\right)
	\mathrm{d}y.
	\label{eq:apd_wij}
\end{equation}
Using the explicit form of the wave functions $\psi_k$ of Fock states, Eq.~\eqref{eq:apd_wij} takes the following closed form \cite{johansson2012qutip}:
\begin{equation} \label{eq:fock_laguerre_ij}
    \begin{aligned}
    W_{kl}(x,p) 
	& =
	\frac{(-1)^k}{\pi}
	\left(
	\sqrt{2}(x+ip)
	\right)^{k-l}
	\\
	& \qquad
	\times
	L_{l}^{(k-l)}
	(2x^2+2p^2)
	\exp(-x^2-p^2),
    \end{aligned}
\end{equation}
where $L^{(\alpha)}_n$ is the generalized Laguerre polynomial.
Note that we have assumed $k\geq l$ in the above expression.
The case $k<l$ is easily deduced from the relation $W_{lk}=W_{kl}^\ast$.

From this, we can express the Wigner function of a quantum state described by a density operator $\hat \rho$ as 
\begin{equation}
	W(x,p) = 
	\sum\limits_{k=0}^{\infty}\sum\limits_{l=0}^{\infty}
	\rho_{kl} \,
	W_{kl}(x,p).
	\label{eq:rho_to_wig}
\end{equation}
Note that $W_{kl}$ takes in general complex values.
However, since $W_{lk} = W_{kl}^{\ast}$, the total Wigner function is real-valued.
In conclusion, it is straightforward to go from any density matrix $\rho_{kl}$ in the Fock basis to its corresponding Wigner function $W$ by using \eqref{eq:rho_to_wig} and \eqref{eq:fock_laguerre_ij}.

In our simulations, we consider mixtures of superpositions of Fock states up to a given photon number $N$, which are described by density operators in a $M$-dimensional Hilbert space (with $M=N+1$).
In order to uniformly sample the set of density operators, we follow the method described in \cite{zyczkowski1998volume}.
We first draw a unitary matrix $U\in\mathbb{C}^{M\times M}$ with matrix elements $u_{ij}$, from which we extract a non-negative normalized vector of eigenvalues $\bm{\lambda} = (\lambda_1, \lambda_2,...,\lambda_M)=\left(\vert{u_{11}}\vert^2, \vert{u_{12}}\vert^2,..., \vert{u_{1M}}\vert^2 \right)$.
We then draw another unitary matrix $V\in\mathbb{C}^{M\times M}$ that we apply to the diagonal matrix defined by $\bm{\lambda}$:
\begin{equation}
	\rho = V
	\begin{pmatrix}
		\lambda_{1} & 			& \\
		& \ddots 	& \\
		& 			& \lambda_{M}
	\end{pmatrix}
	V^{\dagger}.
	\label{eq:random_rho_unitary}
\end{equation}
Both unitary matrices $U$ and $V$ are randomly drawn according to the Haar measure.

\subsection{Discretization}
To perform our numerical simulations, we discretize the functions $W_{kl}$ over a square grid.
We set the size of the grid as $-L/2\leq x \leq L/2$ and $-L/2\leq p\leq L/2$, and we discretize the $x$ and $p$ axes over $n_L$ points that are regularly spaced from $-L/2$ to $L/2$. The parameter $L$ should be chosen in accordance with $N$ (the highest Fock state) such that $\iint_{L^2} W_n(x,p)\approx 1$.
Then, $n_L$ should be chosen such that $n_L\gg L$.
In our simulations, we have generally chosen $N=10$, $L=12$ and $n_L=1000$.

\begin{figure}[t]
	\includegraphics[width=\linewidth]{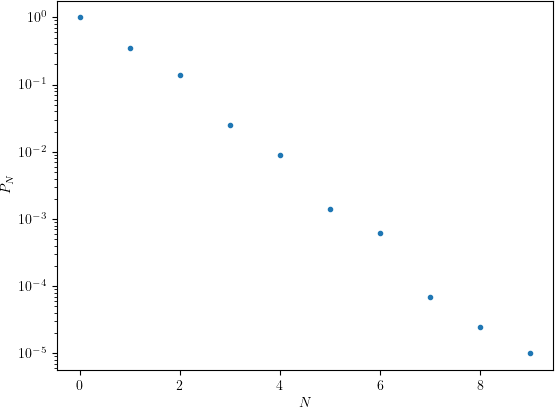}
	\caption{Logarithmic plot of $P_N$ as a function of $N$, where 
	$P_N$ is the probability to get a Wigner-positive state when uniformly sampling the set of density operators containing up to $N$ photons.
	For each value of $N$, we estimate $P_N$ from a set of $10^5$ random states.}
	\label{fig:pn_exp}
\end{figure}

Once we have computed the Wigner function $W$ associated with a random quantum state, we check whether it is positive.
If it is the case, the next step is to compare the cumulative integral of $W$ to the one of vacuum.
The cumulative integral of $W$ can be approximated in the following manner.
We first convert the $n_L\times n_L$ matrix $W$ in a $1\times n_L^2$ vector $\mathbf{v}$.
We then sort the vector by decreasing order which gives us $\mathbf{v}^\downarrow$, and compute its cumulative sum $\mathbf{S}$:
\begin{equation}
	\left(\mathbf{S}\right)_i = 
	\left(\frac{L}{n_L-1}\right)^2
	\ 
	\sum\limits_{k=1}^{i}\mathbf{v}^\downarrow_k.
\end{equation}
The constant $L^2/(n_L-1)^2$ accounts for normalization and corresponds to the surface element associated with a point of the square grid (both ends of the length-$L$ segment are included).
Similarly to $\mathbf{v}$, $\mathbf{S}$ is a $1\times n_L^2$ vector.
Note that the cumulative sum $\mathbf{S}$ is not simply the discretization of the cumulative integral $S_W(s)$ as defined in \eqref{eq:cumulative_integral}.
In the cumulative integral $S_W(s)$, the value of the parameter $s$ defines an area equal to $\pi s^2$, whereas the $i^\text{th}$ component of the cumulative sum $\mathbf{S}$ corresponds to an area of $i\times L^2/(n_L-1)^2$. 
Therefore, $\mathbf{S}$ and $S_W$ are related as:
\begin{equation}
	\left(\mathbf{S}\right)_i 
	\approx 
	S_W\left(\sqrt{\frac{i}{\pi}}\frac{L}{n_L-1}\right).
\end{equation}

Finally, checking numerically that a positive Wigner function $W_1$ majorizes another positive Wigner function $W_2$ amounts to checking that the difference of their respective cumulative sums is positive :
\begin{equation}
	W_1\succ W_2
	\quad
	\Leftrightarrow
	\quad
	\left(\mathbf{S}_1-\mathbf{S}_2\right)_i\geq 0
	\quad\forall i.
\end{equation} 
For simplicity, we introduce the area parameter $a=\pi s^2$ and make the slight abuse of notation to designate by $S_W(a)$ the function $S_W(s=\sqrt{a/\pi})$.

\subsection{Further numerical evidences}

\begin{figure}[t]
	\vspace{2em}
	\includegraphics[width=\linewidth]{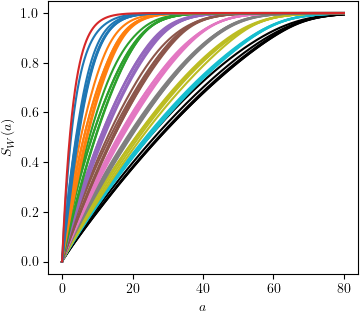}
	\caption{
		 Cumulative integral $S_W(    a)$ of random Wigner-positive states generated according to \eqref{eq:random_rho_unitary} for $N\in\lbrace 0,..., 10\rbrace$.
		For each value of $N$, the cumulative integrals of 5 random Wigner-positive states are plotted.
		Colors are associated to $N$ as follows: red to $N=0$, blue to $N=1$, orange to $N=2$, green to $N=3$, purple to $N=4$, brown to $N=5$, pink to $N=6$, gray to $N=7$, olive to $N=8$, cyan to $N=9$, black to $N=10$.
		We observe that cumulative integrals are in general smaller for high $N$ than for small $N$.
		Also, notice that every state satisfies the conjecture since the cumulative integral of vacuum ($N=0$, red line) is the highest curve.
	}
	\label{fig:rand_cum_int}
\end{figure}

Let us first make an observation on the fraction of Wigner-positive states among the set of quantum states. We call $P_N$ the probability to get a Wigner-positive state when randomly sampling the set of density operators containing up to $N$ photons (we use a similar idea to what is done in~\cite{zyczkowski1998volume}).
As one can see in Figure \ref{fig:pn_exp}, the probability $P_N$ decreases exponentially as a function of $N$. Consequently, it becomes increasingly difficult to randomly generate Wigner-positive states as we expand the size $N$ of the Fock space. For this reason, we have limited our numerical exploration to $N=10$. Figure \ref{fig:rand_cum_int} then illustrates the validity of the conjecture for random Wigner-positive states drawn according to the previously described technique up to $N=10$.

At this point, it is interesting to mention another technique to generate random Wigner-positive states in a deterministic fashion, \textit{i.e.,} with probability $1$ that the random state is Wigner-positive.
In \cite{VanHerstraeten2020}, we highlighted a particular set of Wigner-positive states built with a balanced beam-splitter.
Let us consider a state $\hat{\sigma}$ built from two single-mode states $\hat{\rho}_A$ and $\hat{\rho}_B$ as follows:
\begin{equation}
	\hat{\sigma} =
	\mathrm{Tr}_B
	\left[
	\hat{U}_{1/2}
	\Big(
	\hat{\rho}_A
	\otimes
	\hat{\rho}_B
	\Big)
	\hat{U}^\dagger_{1/2}
	\right],
	\label{eq:sigma_bs}
\end{equation}
where $\hat{U}_{1/2}$ is the two-mode unitary corresponding to a  beam-splitter of transmissivity $1/2$.
Such a state $\hat{\sigma}$ is Wigner-positive for \textit{any} choice of $\hat{\rho}_A$ and $\hat{\rho}_B$ (see \cite{VanHerstraeten2020}).
We can thus use Eq. \eqref{eq:sigma_bs} with random $\hat{\rho}_A$ and $\hat{\rho}_B$ to generate random states that will be Wigner-positive with certainty.
It should be noted, however, that this technique does not span the entire Wigner-positive set since there exist Wigner-positive states that cannot be expressed in the form of \eqref{eq:sigma_bs}, see \cite{VanHerstraeten2020}.

\begin{figure}
	\vspace{2em}
	\includegraphics[width=\linewidth]{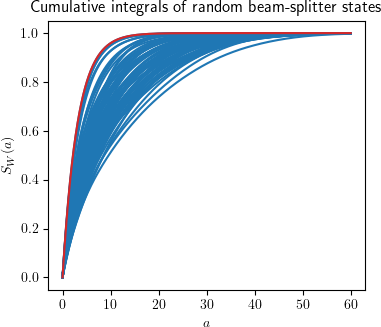}
	\caption{
		Cumulative integral $S_W(    a)$ of random (Wigner-positive) beam-splitter states generated according to \eqref{eq:sigma_bs}.
		We choose the inputs of the beam-splitter $\hat{\rho}_A$ and $\hat{\rho}_B$ to be pure states in the form of \eqref{eq:limited_superposition_fock}, with respectively up to $N_A$ and $N_B$ photons.
		Each blue line corresponds to a random instance of beam-splitter state for some $N_A$ and $N_B$ such that $N_A+N_B\leq 10$.
		The red line is the cumulative integral for the vacuum state; it is greater than any blue line as expected from our conjecture.
	}
	\label{fig:rand_cum_int_bs}
\end{figure}

In \eqref{eq:sigma_bs}, notice that if $\hat{\rho}_A$ and $\hat{\rho}_B$ are mixed states, $\hat{\sigma}$ can be decomposed as a mixture of beam-splitter states built from pure states.
To test our conjecture over the set of beam-splitter states, it is thus sufficient to limit our study to $\hat{\rho}_A,\hat{\rho}_B$ being pure states (see majorization property \eqref{eq:majorization_convex_mixture}).
Let us consider pure states $\ket{\psi_i}$ which are finite superpositions of the first Fock states:
\begin{equation}
	\ket{\psi_i}
	=
	\sum\limits_{k=0}^{N_i}
	c^{(i)}_k\ket{k}
	\label{eq:limited_superposition_fock}
\end{equation}
where $i\in\lbrace A,B\rbrace$.
With $\ket{\psi_A}$ and $\ket{\psi_B}$ containing respectively up to $N_A$ and $N_B$ photons, the state $\hat{\sigma}$ belongs to the span of Fock states up to $N=N_A+N_B$.
We have tested the conjecture for various choices of $(N_A, N_B)$ such that $N_A+N_B\leq 10$, and have found that each instance satisfies the conjecture.
This is illustrated in Figure \ref{fig:rand_cum_int_bs}.

\bibliographystyle{linksen}
\bibliography{contmaj}

\end{document}